\documentclass[12pt]{article}

\usepackage{graphicx}
\begin{document}
                   
\begin{center}
{\Large \bf
Variational approach for walking solitons in birefringent fibers}

\vspace{0.8cm} 
\renewcommand{\thefootnote}{\fnsymbol{footnote}}
\setcounter{footnote}{1}
{\bf {N.~J.~Rodr\'{\i}guez-Fern\'andez \footnote{Present address:
Observatorio Astr\'onomico Nacional, 
Apdo. 1143, Alcal\'a de Henares 28800, Spain} }
and J.~M.~Soto-Crespo} 
\\
\vspace{0.1cm}
Instituto de \'Optica, Consejo Superior de Investigaciones
Cient\'{\i}ficas \\ Serrano 121, 28006 Madrid, Spain \\
\vspace{0.5cm}

{\bf ABSTRACT}
\end{center}
\vspace{0.1cm}

\oddsidemargin 1cm
\topmargin 0 cm
\textheight 22.5 cm
\textwidth 16 cm
We use the variational method to obtain approximate analytical expressions
for the
 stationary pulselike solutions in 
birefringent fibers when differences in
both phase velocities and group velocities between the two components
and rapidly oscillating terms
are taken into account.  After checking the validity of the approximation
we study how the soliton pulse shape depends on its velocity and nonlinear 
propagation constant. By numerically solving the propagation equation
we have found that most of these stationary solutions are stable.

\newpage
\section{Introduction}
\label{intro}
Pulse propagation in nonlinear birefringent optical fibers is
presently an active area of research 
[1-12].Even ``single-mode"  fibers support two degenerate 
modes polarized  in two orthogonal directions. The degeneracy can
be broken introducing deliberately
a large amount of birefringence
through design modifications giving place to the so called
polarization-preserving fibers.   Weak light pulses propagating
in these media maintain its state of polarization when they are initially
polarized along either of the principal axes. However at high
intensities nonlinear birefringence can change drastically
the propagation dynamics.  Birefringence can also be a residual
effect due to imperfections in the manufacturing process.

	The equations that govern 
pulse  propagation in nonlinear birefringent pulses 
have been derived by
Menyuk \cite{menyuk}.  Two main cases of 
birefringence: i) high and ii) low, 
have been usually considered separately.  In the approximation
of high birefringence \cite{menyuk,pare},  phase and group
velocities of both field components are considered to be different.
At the same time rapidly oscillating nonlinear terms are neglected.
On the other hand, the approximation of low birefringence
takes into account the difference in phase velocities between
the two linearly polarized components, but neglects their 
difference in group velocities. The full polarization dynamics
of solitons in this last
approximation has been studied in \cite{nai1}.

	The general case has been analyzed in just a few
papers \cite{tor,my1,evan}.
New vectorial stationary solitons for this case were 
first discovered in \cite{my1},
and later on the whole family of stationary pulselike
solutions was numerically determined
by Torner et al. \cite{tor}.  These families of solitons
correspond to mutually trapped pulses that
propagate with a common
group velocity and which therefore can be
called walking vector solitons \cite{to2}.
Stationary solutions as singular points of this
infinite dimensional dynamical system play a pivotal role
to understand the  propagation dynamics of any arbitrary
input pulse. Here we generalize the method described in \cite{pare}
to find accurate analytical
approximations of the solitary wave solutions for
the case of high birefringence.
We generalize it for the more general case, when no terms are neglected.
Our approximate results are  compared with existing
exact ones \cite{tor,my1}
 to  prove the validity of our approach. 
 We then examine the behavior of the two pulse components
 when  the  propagation constant ($q$) or the velocity~($v$) change.
We find that  the width of the pulses is determined by $q$ whereas
 the amplitudes ratio $a_1/a_2$ is determined by $v$ when the energy is higher
than a certain value.
By means of
numerical propagation of the stationary solutions found in this way,
we observe that most of them are stable. 

	The rest of the paper is organized as follows. 
The problem to be addressed
by the variational method is described in section
\ref{sta}. Our variational approach is developed
in section \ref{var}. In section \ref{num} we present
the numerical results and finally  in section \ref{sum}
we briefly summarize the  main conclusions.

\section{Statement of the problem}
\label{sta}

Pulse propagation in a birefringent optical fiber
can be described in terms of two nonlinearly coupled nonlinear
Schr\"odinger equations. In a reference frame traveling 
along the $\xi$ axis with the average group velocity,
and in normalized units, this set takes the form~ \cite{menyuk,evan}
\begin{eqnarray}
\label{propa}
& &- i U_{\xi}= i \delta U_{\tau} +\frac{1}{2} U_{\tau\tau}
+\left( |U|^2+ A |V|^2 \right) U
+\beta U +B V^2 U^*, \nonumber\\
& & -i V_{\xi}=-i \delta V_{\tau} +\frac{1}{2} V_{\tau\tau}
+\left( |V|^2+ A |U|^2\right) V  -\beta V + B U^2 V^* ,
\end{eqnarray}
where $U$ and $V$ are the slowly varying envelopes of the two
linearly components of the field along
the $x$ and $y$ axis respectively, $\xi$ is the normalized propagation
coordinate, $\delta$ is the inverse group velocity difference, 
$\beta$ is half the difference between the propagation constants, 
$A$ is the normalized ratio of the nonlinear susceptibilities, 
$B=1-A$, $\tau$ is the
normalized retarded time, and the asterisk denotes complex conjugation.
In linearly birefringent fibers $A=2/3$, as we set here. The
set of Eqs.\ref{propa}
has at least three integrals of motion \cite{my1}, i) the
action (total energy), ii) the momentum and iii)
the Hamiltonian,
which are a consequence of i)the translational invariance of
Eqs.(\ref{propa}) relative to phase shifts, ii) invariance in $\tau$ and
iii) translational invariance in $\xi$.

	Eqs.\ref{propa} have two simple linearly polarized pulselike
solutions,
viz., linearly polarized soliton waves along the slow axis
\begin{equation}
\label{slow}
U=\frac{\sqrt{2(q-\beta)}}{\cosh{[\sqrt{2(q-\beta)}(\tau-\delta\xi)]}}
\exp{(i\, q \, \xi)},\;\;\; V=0
\end{equation}

and linearly polarized soliton waves along the fast axis
\begin{equation}
\label{fast}
U=0,\;\;V=\frac{\sqrt{2(q+\beta)}}{\cosh{[\sqrt{2(q+\beta)}(\tau+\delta\xi)]}}
\exp{(i \,q \, \xi)}.
\end{equation}
Their corresponding values for the total energy of the pulse
are:
\begin{equation}
\label{QU}
Q=\int_{-\infty}^{\infty} (|U|^2+|V|^2) d\tau= 2\sqrt{2(q\pm\beta)}
\end{equation}
	The stability of these solutions has been determined
in Ref.\cite{nai1} for $\delta=0$, and for arbitrary 
nonzero values
of $\delta$ in Ref.\cite{my1}.

	Usually  Eqs.\ref{propa} are written 
without the last two  terms \cite{menyuk},  as terms with 
coefficient $\beta$ can be eliminated
by $z$ dependent phase transformations of $U$ and $V$, and this in turn
gives the last nonlinear terms rapid phase variations with $z$ which
in principle allow their neglect. We must however 
bear in mind that the last
terms in Eq.\ref{propa} are the only ones responsible for energy transfer
between both polarizations. And although averaging out the fast oscillatory
terms has proven to be a good approximation to describe most
of the observed phenomena \cite{barad,kutz,chbat} in the picosecond regime,
we will go further in
retaining these terms in the analysis.

\section{Variational approach}
\label{var}
	The variational approach when applied to
the single Nonlinear Schr\"odinger Equation (NLS) was introduced
by Anderson \cite{anderson}. Since then it has been
widely used for coupled NLS equations. For the specific case of
birefringent fibers in the high birefringence
approximation it has been used for studying 
dynamical behaviors \cite{kutz,ued,anderson2,kaup,wang},
as well as the stationary case \cite{pare}. 

	We look for stationary solutions moving
at a common velocity $v$ in our frame of reference.
They can be written \cite{tor} as
\begin{eqnarray}
\label{asu}
&& U(\tau ',\xi)=P_1(\tau ')\exp{(i \,q\, \xi)} \nonumber
\\
&&  V(\tau ',\xi)=P_2(\tau ')\exp{(i \,q\, \xi)}
\end{eqnarray}
where $\tau '=\tau- v\xi$ is the common retarded time. 
Inserting Eqs.~\ref{asu}
into Eqs.\ref{propa}, we get
a set of ordinary differential equations (ODE's) for $P_{1,2}$, 
which reads
\begin{eqnarray}
\label{propan}
&& i (\delta-v) \dot P_1 +(\beta-q) P_1+\frac{1}{2}\ddot P_1
+(|P_1|^2+ A|P_2|^2)P_1+BP_2^2 P_{1}^{*}=0
\nonumber \\
&&-i (\delta+v) \dot P_2 -(\beta+q) P_2+\frac{1}{2}\ddot P_2
+(|P_2|^2+ A|P_1|^2)P_2+BP_1^2 P_{2}^{*}=0
\end{eqnarray}
where the overdots indicate derivative respect to $\tau '$.
Eqs. (\ref{propan}) can be derived (via the Euler-Lagrange 
equations) from the following Lagrangian 
$L=\int_{-\infty}^{\infty} {\cal L} d\tau'$, where
the lagrangian density $\cal L$ is given by
\begin{eqnarray}
\label{lagran}
&{\cal L}=&|\dot P_1|^2+|\dot P_2|^2 
-i(\delta-v)\left(P_1^{*}\dot P_1-P_1\dot P_1^{*}\right)
\nonumber \\
& & 
-i(\delta+v)\left(-P_2^{*}\dot P_2+P_2\dot P_2^{*}\right)
+2 (q-\beta)|P_1|^2
\nonumber \\
& &
+2 (q+\beta)|P_2|^2
-|P_1|^4-|P_2|^4
-2  A |P_1|^2 |P_2|^2
\nonumber \\
& &
-B \left(P_1^2P_2^{*\,2}+P_1^{2\,*} P_2^{2}\right).
\end{eqnarray}
	
We now assume the following ansatz for $P_{1,2}$
\begin{eqnarray}
\label{ansatz}
& P_{i}=a_{i}\; sech(b_{i}\,\tau ')\exp{(i\,c_{i}\tau ')},
&\;\;\;\;\;\; i=1,2.
\end{eqnarray}
Where the variational
parameters $a_{1,2}$, $b_{1,2}$, and $c_{1,2}$ are assumed 
to  have real values.
The above ansatz is inspired from solutions \ref{slow}, and \ref{fast},
as well as
from the work of Par\'{e} \cite{pare}. We know that the exact
solution has a complex phase chirp \cite{tor} induced
by the four wave mixing term (last term in Eqs.\ref{propa}), 
which we neglect for simplicity but also because
from previous numerical calculations we observed  that the phase
dependence on $\tau '$ was mainly lineal
\cite{my1}. 
	
By introducing Eqs.\ref{ansatz} into Eqs.\ref{lagran},
and applying the Euler-Lagrange equations to the averaged
Lagrangian one obtains, after some straightforward
algebra, a set of six  nonlinear  coupled
equations,viz,
\begin{equation}
\label{sis1}
c_1= -\delta+v +B\,b_1a_2^2 I(c_-)
\end{equation}
\begin{equation}
\label{sis2}
c_2= \delta+v -B\,b_2\,a_1^2  I(c_-)
\end{equation}
\begin{equation}
\label{sis3}
\frac{b_1}{3}+\frac{2D_1}{b_1}-\frac{4 a_1^2}{3b_1}-
A\frac{a_2^2}{b_2}m_0(\eta)-B\,a_2^2\, M(b_1,b_2,c_-)=0
\end{equation}
\begin{equation}
\label{sis4}
\frac{b_1}{3}-\frac{2D_1}{b_1}+\frac{2 a_1^2}{3b_1}-
A\frac{a_2^2\eta}{b_2}\dot m_0(\eta)-B\, 
a_2^2b_1 \frac{\partial}{\partial b_1}
 M(b_1,b_2,c_-)=0
\end{equation}
\begin{equation}
\label{sis5}
\frac{b_2}{3}+\frac{2D_2}{b_2}-\frac{4 a_2^2}{3b_2}-
A\frac{a_1^2}{b_1}m_0(\eta^{-1})-B\,a_1^2\, M(b_1,b_2,c_-)=0
\end{equation}
\begin{equation}
\label{sis6}
\frac{b_2}{3}-\frac{2D_2}{b_2}+\frac{2 a_2^2}{3b_2}-
A\frac{a_1^2}{b_1\eta}\dot m_0(\eta^{-1})-B\,
a_1^2b_2 \frac{\partial}{\partial b_2}
 M(b_1,b_2,c_-)=0
\end{equation}

where 
\begin{equation}
\label{icm}
I(x)=\int_{-\infty}^{\infty}
\tau \sin{(2x \tau)} sech^2(b_1\tau)sech^2(b_2\tau) d\tau
\end{equation}
\begin{equation}
\label{cemn}
c_-=c_2-c_1=2\delta-\frac{b_1b_2}{2} Q B I(c_-)
\end{equation}
\begin{equation}
\label{eme}
m(x)=\int_{-\infty}^{\infty} sech^2 t\,sech^2(x\,t) dt
\end {equation}
\begin{equation}
\label{eMe}
M(b_1,b_2,c_-)=
\int_{-\infty}^{\infty} cos(2c_-t) sech^2(b_1t) sech^2(b_2 t) dt.
\end{equation}
$D_1=q-\beta+c_1^2/2+(\delta-v)c_1$,
$D_2=q+\beta+c_2^2/2-(\delta+v)c_2$, $\eta=b_1/b_2$, and
$Q=2(a_1^2/b_1+a_2^2/b_2)$

In order to simplify the set of equations \ref{sis1} - \ref{sis6}, we
need to do some assumptions and approximations.
We first take into account \cite{tor,my1} that the main
contribution to $c_-$  becomes from $2 \delta$, while the integral is 
a minor correction. Therefore, using 
the first order B\"orn approximation \cite{morse}
$c_-$ becomes 
\begin{eqnarray}
\label{cminus}
& &c_-\approx2\delta+\frac{b_1b_2}{2} Q B I(2 \delta)
\end{eqnarray}
 Introducing
Eq.\ref{cminus} into Eqs.\ref{sis3} - \ref{sis6} we get a reduced set
of four nonlinear equations. The next step must consist
in finding
accurate analytical expressions
for the integrals $I$, $m$, and $M$ which
can not be exactly integrated.
Following ref.\cite{pare} we expand $m$ in a Taylor series around $\eta=1$
up to second order, and obtain:
\begin{eqnarray}
\label{emeap}
&&m(x)=\frac{4}{3}+\frac{2}{3}(1-x)+\frac{2}{3}(1-\frac{\pi^2}{15})(1-x)^2
\nonumber\\
&& \dot m(x)=-\frac{2}{3}-\frac{4}{3}(1-\frac{\pi^2}{15})
(1-x)+(\frac{\pi^2-10}{5})(1-x)^2
\end{eqnarray} 
	The same Taylor expansion for $M$ or $I$ gives  much more
complicated expressions in terms of Polygamma functions 
\cite{abram}, which should be evaluated numerically. Instead
of this approach we choose
to approximate $M$ and $I$ by replacing $b_i$ by $(b_1+b_2)/2$. 
In this way
the integrals can be analytically performed to give:
\begin{eqnarray}
\label{eMeap}
& & M\approx \frac{16\pi c_-}{3b^2 }\left(1+
\left(\frac{2c_-}{b}\right)^2\right) 
csch\left[\frac{2 \pi c_-}{b}\right] 
\\ && \nonumber \\
& &  \frac{\partial M}{\partial b_i}= 
\frac{-108 \pi c_-^3}{3b^5}
csch\left[\frac{2 \pi c_-}{b}\right]
-\frac{32 \pi c_-[b^2+4 c_-^2] }{3 b^5} 
csch \left[\frac{2 \pi c_-}{b}
\right] \nonumber \\ & &\nonumber \\
& &   +\frac{32 \pi^2 c_-^2[b^2+4 c_-^2]}{3 b^6}
coth\left[\frac{2 \pi c_-}{b}\right] 
csch\left[\frac{2 \pi c_-}{b}\right] 
\end{eqnarray}
\begin{eqnarray}
\label{cminusap}
I(x)
\approx
\frac{16 \pi x [b^2+x^2]}{3b^5}
\left[\pi coth(\frac{2 \pi x}{b}) -\frac{b}{2x}
-\frac{4 b x}{b^2+4 x^2}\right] 
csch[\frac{2 \pi x}{b}],
\end{eqnarray}
where $b=b_1+b_2$.

	From here the procedure goes as follow. Given
the equation coefficients $\delta$, $\beta$, $A(=2/3)$,
and $B(=1/3)$, together with the parameters $v$ and $q$,
we solve the Eqs.\ref{sis3} -\ref{sis6} to obtain
$b_{1,2}$, and $a_{1,2}$.  Then $c_{1,2}$ are
obtained from Eqs.\ref{sis1}- \ref{sis2}.

\section{Numerical results}
\label{num}
	We have numerically solved the four coupled
nonlinear algebraic equations by a Powell hybrid method
\cite{powell}.
In all cases a unique solution was instantaneoulsy
obtained. In order to
check the validity of the
approximations used to solve the integrals we sometimes
evaluated numerically $M$, $m$, and $I$  at each step getting
almost identical
results but with a much higher CPU time consumption.

\begin{figure}
\includegraphics*[bb=83 154 502 502]{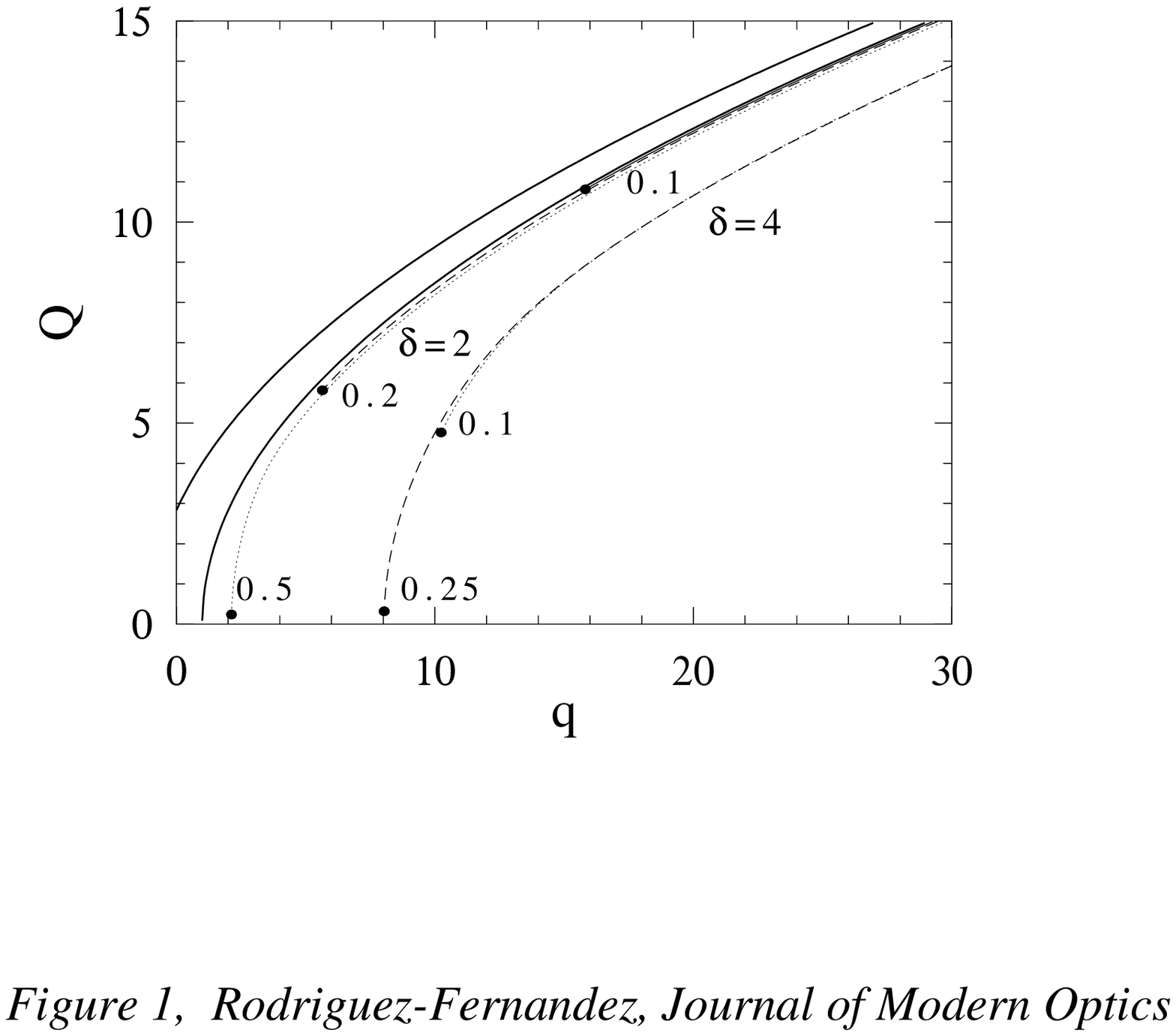}
\caption{ Dependence of the soliton energy $Q$ on the nonliner
propagation constant $q$ for the fast and slow
linearly polarized solitons (continuous lines) and coupled soliton
states for $\beta=1$, $\delta=2$ and $4$, and several values of
the parameter $v$, which is written close to the point
where the curves emerge.
This point is marked with a filled circle}
\label{abu} 
\end{figure}

	Fig.\ref{abu} shows the dependence of the soliton energy~$Q$ 
on the nonlinear propagation constant~$q$. This
diagram was obtained by means of the above described
variational  method.
The solid lines represent the solutions given by
Eqs. \ref{QU}.  Different sets of 
almost parallel curves correspond to
different values of the parameter $\delta$
whilst the value of $v$ hardly influences the shape
of these curves, fixing solely the minimum allowed value
of $q$. These curves coincide totally  with those shown
in fig.1(a) of Ref.\cite{my1}, which were obtained exactly
by numerically  solving the propagation equation for certain
values of $q$ whatever $v$ was. 

In addition to the solutions approximated by Eq.\ref{ansatz},
where
$a_{1,2}$ were assumed to be real, and therefore represent
solutions where $U$ and $V$ are in phase, there exist 
solutions where $U$ and $V$ are $\pi/2$ out of phase \cite{tor}.
They can be obtained directly
from our variational approach
if we do not imposed $a_{1,2}$ to be real. Alternatively
our approach can account for them just by changing the 
ansatz (Eq.\ref{ansatz} for $P_i$). If we choose the following one:
\begin{eqnarray}
\label{ansatz2}
& &\tilde P_1=a_1 sech (b_1\,\tau')exp(i\,c_{1}\tau ')
\nonumber \\
& & \tilde P_2=a_2 sech (b_2\,\tau')exp(i(c_{2} \tau ' +\pi/2)),
\end{eqnarray}
and repeat the process, it is easily realized that the only
change produced in the nonlinear
algebraic equations is to replace $B$ by $-B$. This indicates
that the existence of these independent solutions is intimately related
to the inclusion of the four wave mixing term in the
propagation equations.

\begin{figure}
\includegraphics*[bb=69 130 490 665]{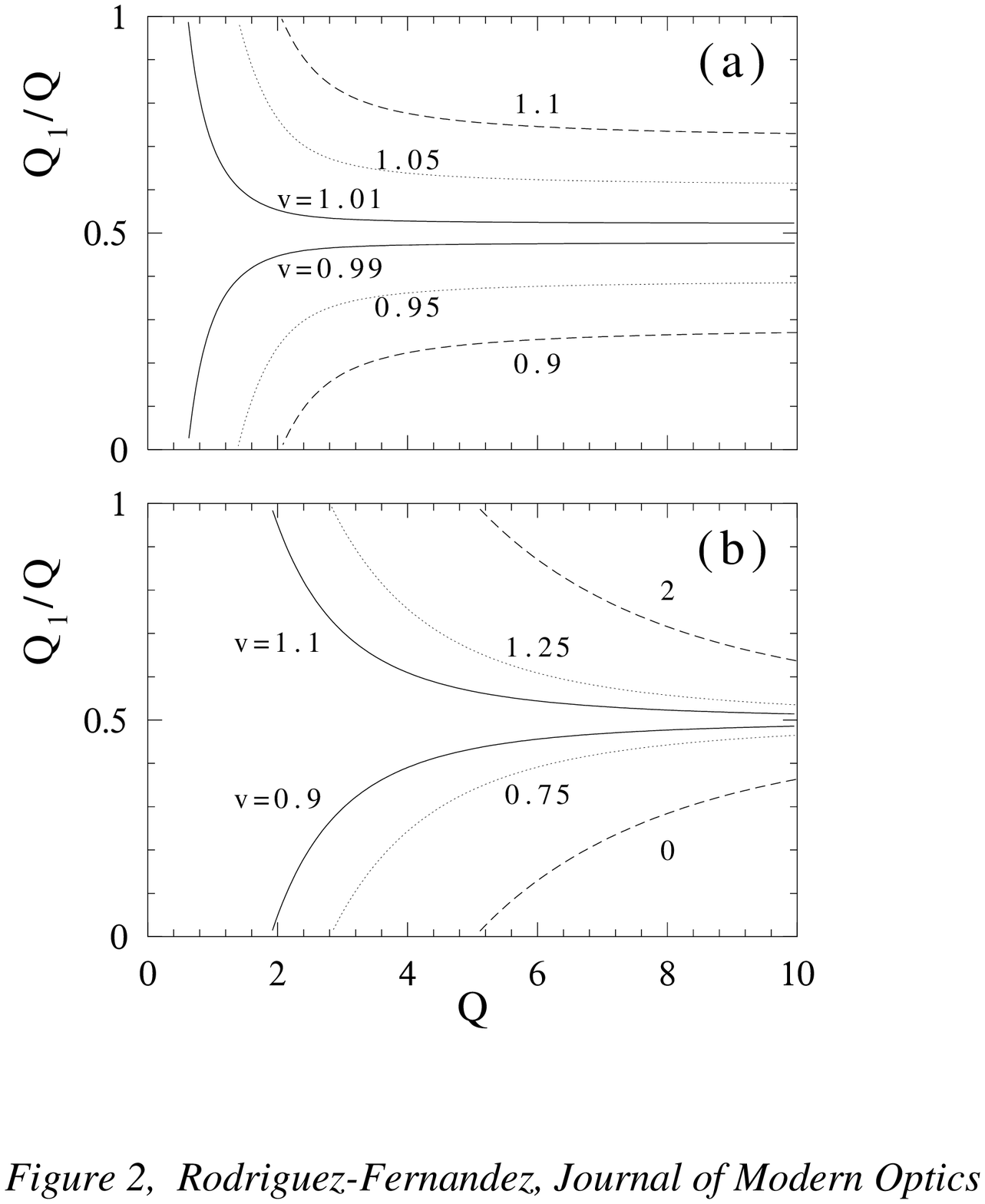} 
\caption{Fraction of energy carried by the walking soliton in the
slow polarization vs. the total energy of the solution.
(a) $U$ and $V$ are in phase. (b) $U$ and $V$ have a relative
phase difference of $\pi/2$. They correspond to 
$\delta=\beta=1$, and different values of $v$, which is written
in the figure.}
\label{ab2} 
\end{figure}

  Figure \ref{ab2} shows  the fraction of energy in the slow
mode ($Q_1/Q$) versus the total energy~($Q$) of the 
two-parameter  family  of solutions  for $\delta=\beta=1$
as obtained from our variational method.  In (a) and (b)
$P_i$ is given by Eq.\ref{ansatz} and  Eq.~\ref{ansatz2}
respectively. The results can be compared with the exact ones
in Fig.~1 of Ref.\cite{tor}, which were found numerically.
The symmetry of these
curves
is a result of the symmetry that possesses the
Lagrangian under:
\begin{eqnarray}
\label{symm}
&& q \longrightarrow \tilde q=q+\frac{ \tilde v^2-v^2}{2},
\;\;\;\;\;
 v  \longrightarrow \tilde v=\frac{2\beta}{\delta} -v \;,
 \nonumber \\
&&c_{2,1} \longrightarrow \tilde c_{2,1}=\frac{2\beta}{\delta} -c_{1,2},\,
\;\;\;\;\;
a_{2} \longleftrightarrow a_{1},\;\;\;\;  b_{2} 
\longleftrightarrow b_{1}
\end{eqnarray}

	Figs.~\ref{abu}, and \ref{ab2} illustrate the
accuracy of the variational approach as developed
in the section \ref{var}. We must remark that
solving the set of four nonlinear algebraic 
equations is an easy task that can be
done really fast. In our computer
(Alpha Dec 2100/500) we obtain thousands of 
these variational solutions
in just a few CPU seconds.

\begin{figure}
\includegraphics*[bb=83 119 449 581]{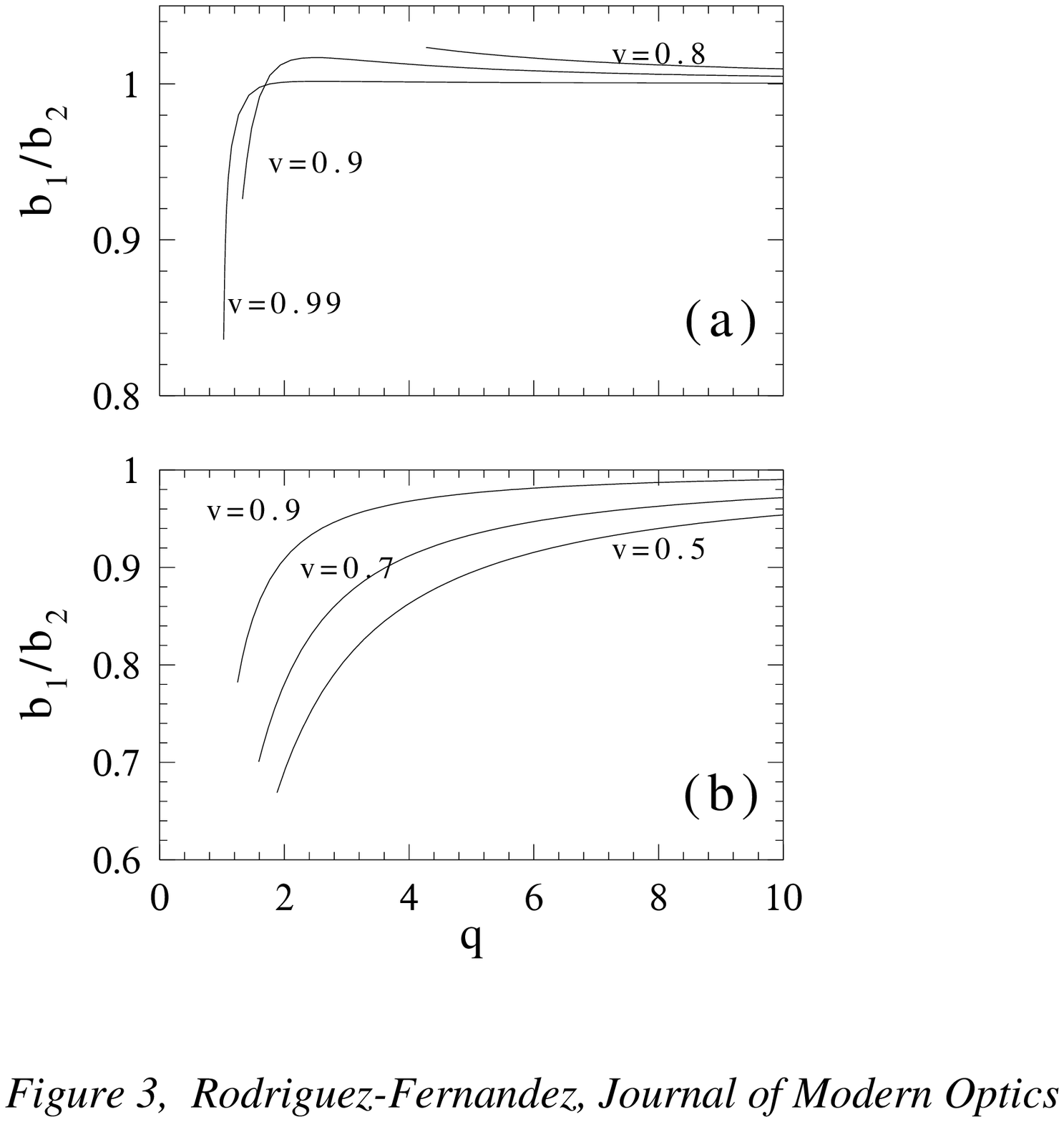} 
\caption{Width ratio ($\eta$) vs $q$  for $\delta=\beta=1$, and
three values of the parameter $v$.
a) $U$ and $V$ are in phase. b) $U$ and $V$ are in quadrature.  }
\label{ab3} 
\end{figure}

	Fig.~\ref{ab3} shows the variation of the
width ratio ($\eta=b_1/b_2$)
vs. the propagation constant $q$ for the stationary solutions 
when $U$ and $V$
are (a) in phase, and (b) 
in quadrature. We take $\beta=\delta=1$, and three values of
$v$ are considered,  which 
are written close to the corresponding curves.  In all cases
$\eta$ tends to $1$ as $q$ increases, i.e. 
for high values of $Q$ (see Fig.~\ref{abu}).
The departure of $\eta$ from unity  is higher when $U$ and $V$ are
in quadrature than when they are in phase. In any case it is small
except around the minimum allowed value of $q$, where the curves emerge and
one of the components ($V$ in the cases of figure~\ref{ab3}) is almost zero.
This figure serves also to verify  {\it a posteriori} that our approximations of 
integrals $I$, $m$ and $M$ had good basis. 

\begin{figure}
\includegraphics*[bb=90 149 452 623]{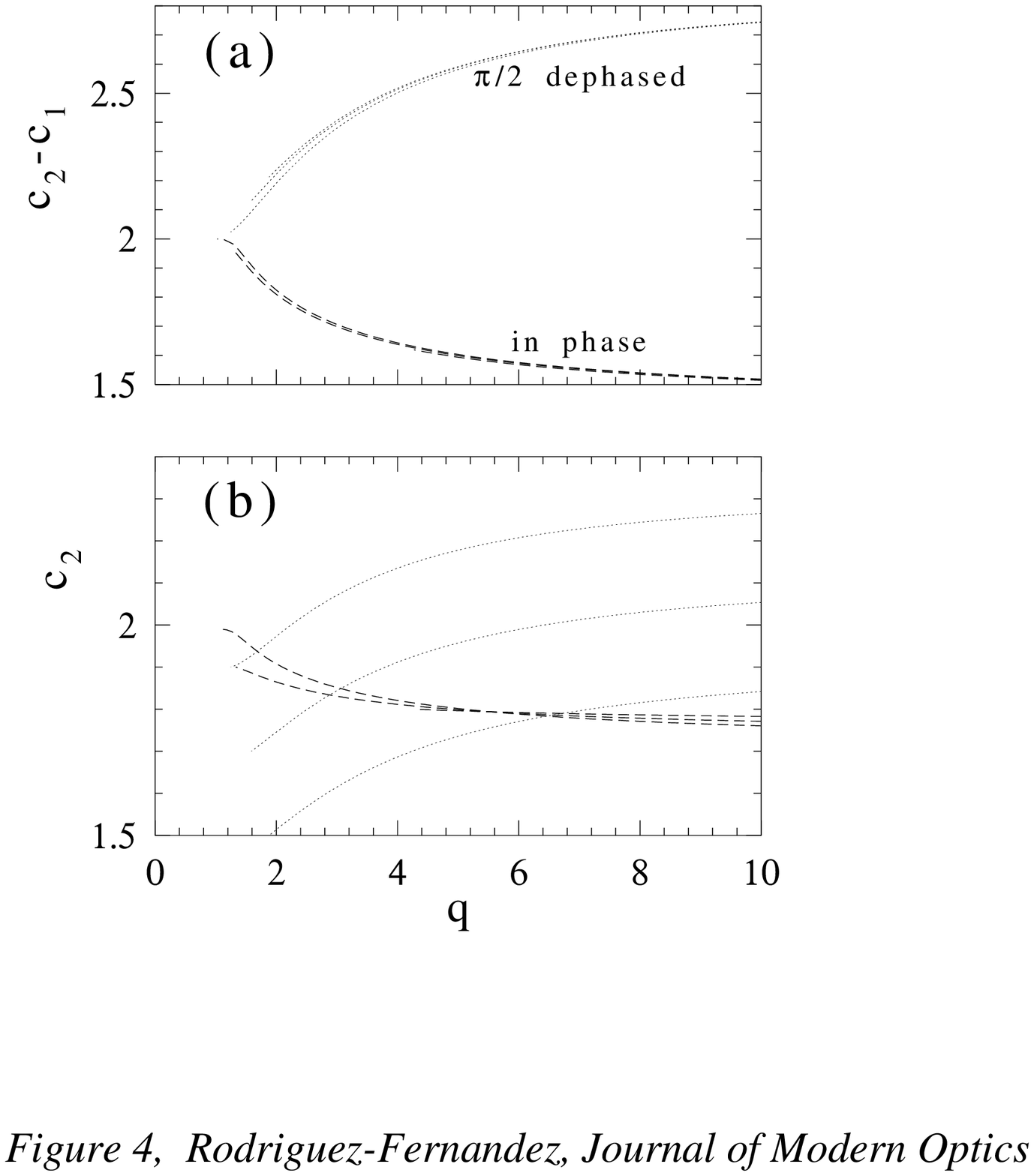} 
\caption{a) Central frequency difference between the soliton
components $V$ and
$U$ vs. $q$. b) Frequency shift of the fast component as a function 
of the propagation constant $q$. The dashed lines represent the
solutions where $U$ and $V$ are in phase and the dotted ones
those in quadrature. The values of the parameters are the same than
in Fig.~\protect \ref{ab3}.}
\label{ab4} 
\end{figure}

Similarly Fig.\ref{ab4}
shows (a) $c_2-c_1$  and (b) $c_2$ vs. $q$  
for the same values of the parameters.
As expected, at the minimum value allowed for $q$, where
the solution is a fast soliton (Eq.\ref{fast}),
$c_2$ is exactly $\delta +v$. As it occurred for $Q$ vs. $q$, the
value of $c_-$ hardly depends on $v$, being mainly determined
by the value of $q$. It is also remarkable that while
the solutions for $U$ and $V$ in phase decrease its frequency difference 
(i.e. $c_-$)
as we increase $q$ (and therefore $Q$), the opposite happens
for the solutions with $U$ and $V$ in quadrature. In these last cases
$c_-$ becomes more
different to    $2\delta$ than when $U$ and $V$ are in phase.
The same happened with $\eta$ respect to $1$ (see Fig.~\ref{ab3}).
In fact we observe that the variational
solutions were much more accurate for the solutions
with $U$ and $V$ in phase than for those in quadrature.

\begin{figure}
\includegraphics*[bb=85 124 416 759, width=8cm]{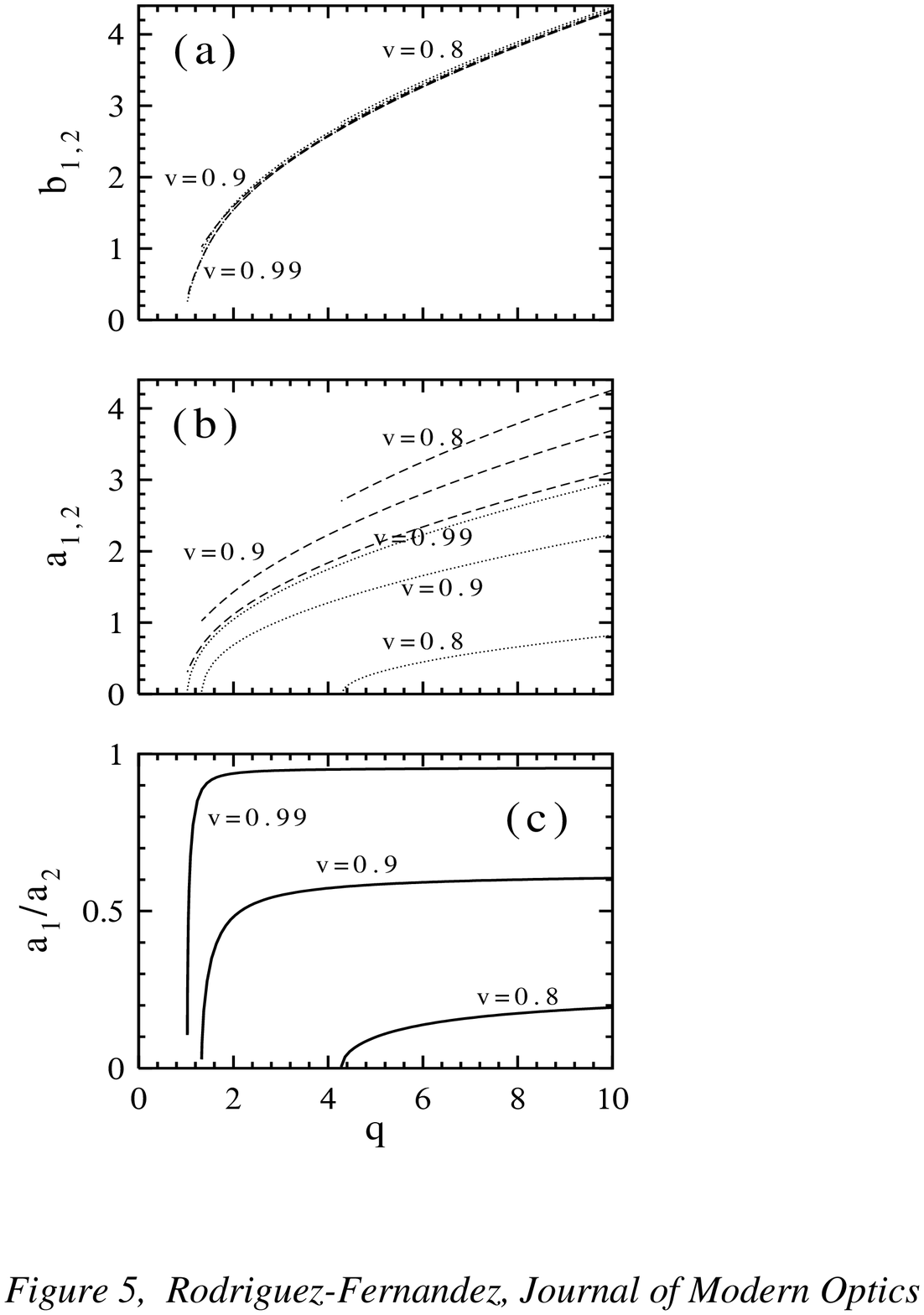} 
\caption{ a) Inverse widths ($b_{1,2}$),  b) peak 
amplitudes ($a_{1,2}$) and c) amplitude ratio ($a_1/a_2$) of the
variational solution for $U$ and $V$ in phase  vs. $q$.
In (a) and (b) the dotted lines are for $U$ and the dotted lines for $V$. 
$\delta=\beta=1$ and the values of $v$ are  written in the figures}
\label{ab5} 
\end{figure}

	Fig.\ref{ab5} shows the variation of the widths and
peak amplitudes vs. $q$ for the solutions with $U$ and $V$ in
phase. It can be seen that for a given value of $\delta$
and $\beta$, the widths depend almost exclusively
on $q$, being almost completely independent on $v$. On the other
hand the asymptotic behavior of $a_1/a_2$ does not depend
on $q$ but only on $v$ (see  Fig.\ref{ab5}c).

\begin{figure}
\includegraphics*[bb=83 154 502 602]{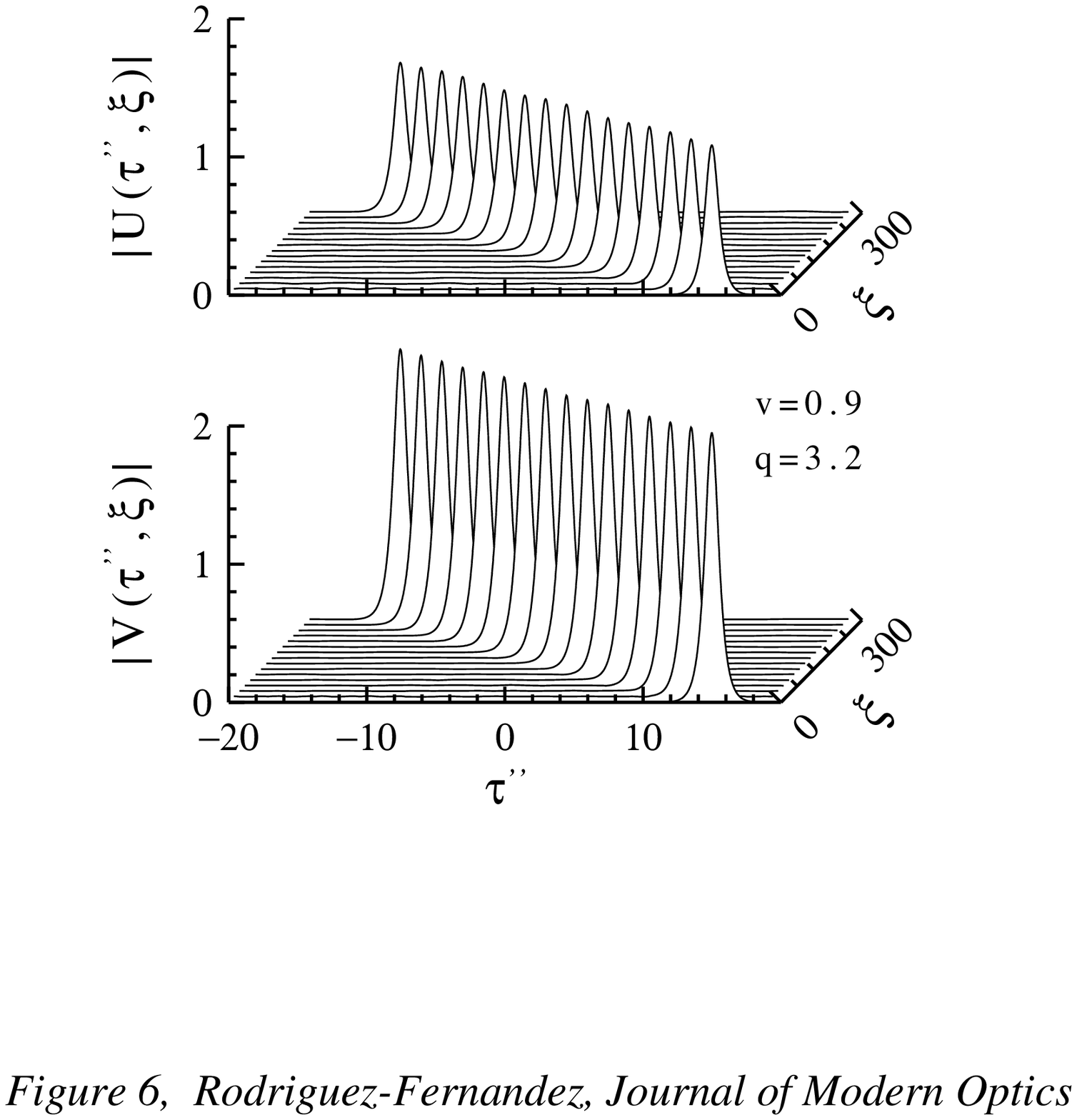} 
\caption{ Stable propagation of a walking soliton with its components
$U$ and $V$ in phase. $\beta=\delta=1$, $q=3.2$, and $v=0.9$ }
\label{ab6}
\end{figure}

\begin{figure}
\includegraphics*[bb=124 170 497 586]{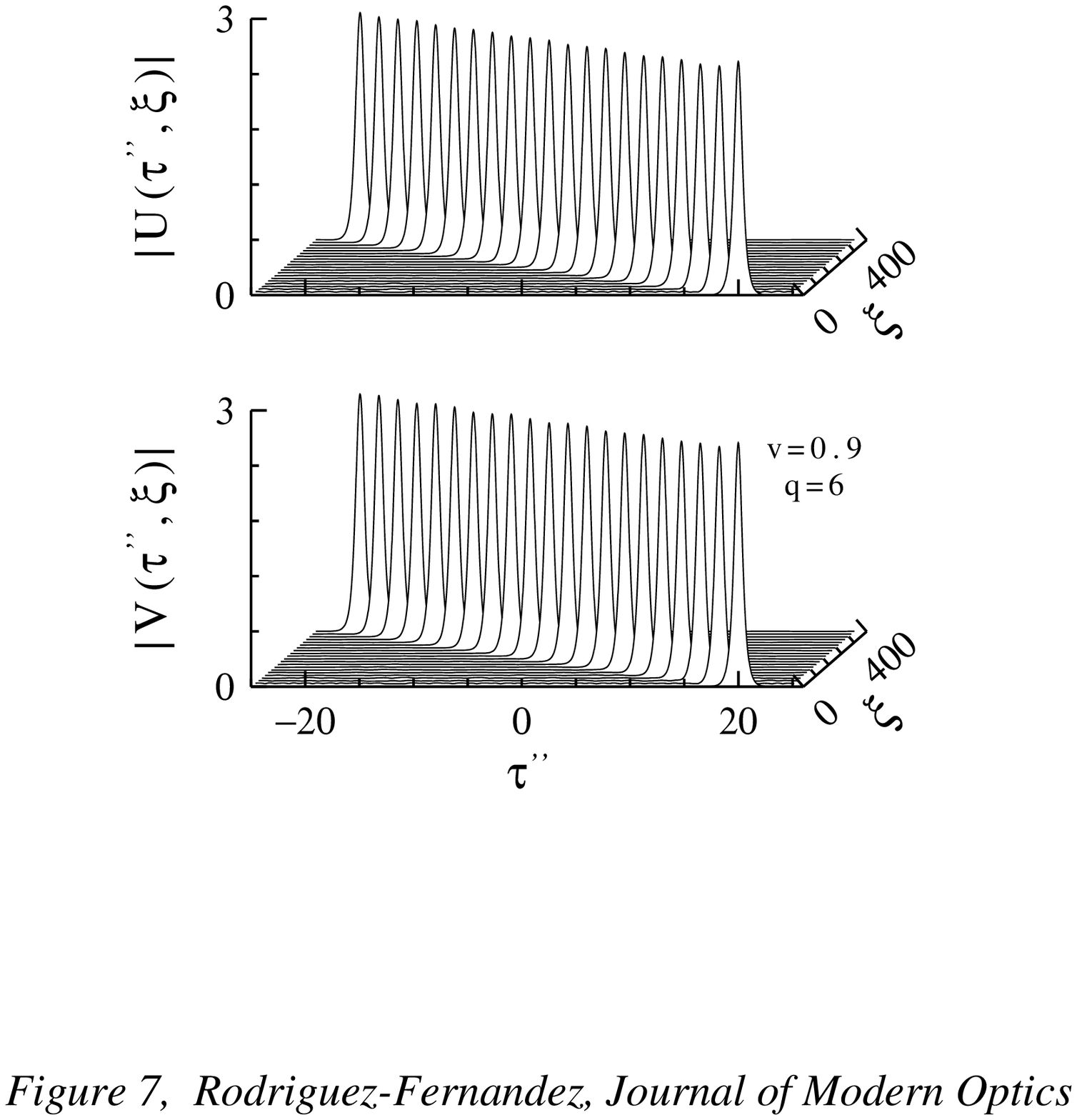} 
\caption{ Stable propagation of a walking soliton with its components
$U$ and $V$ in quadrature.  $\beta=\delta=1$, $q=6$, and $v=0.9$ }
\label{ab7}
\end{figure}

\begin{figure}
\includegraphics*[bb=119 126 512 541]{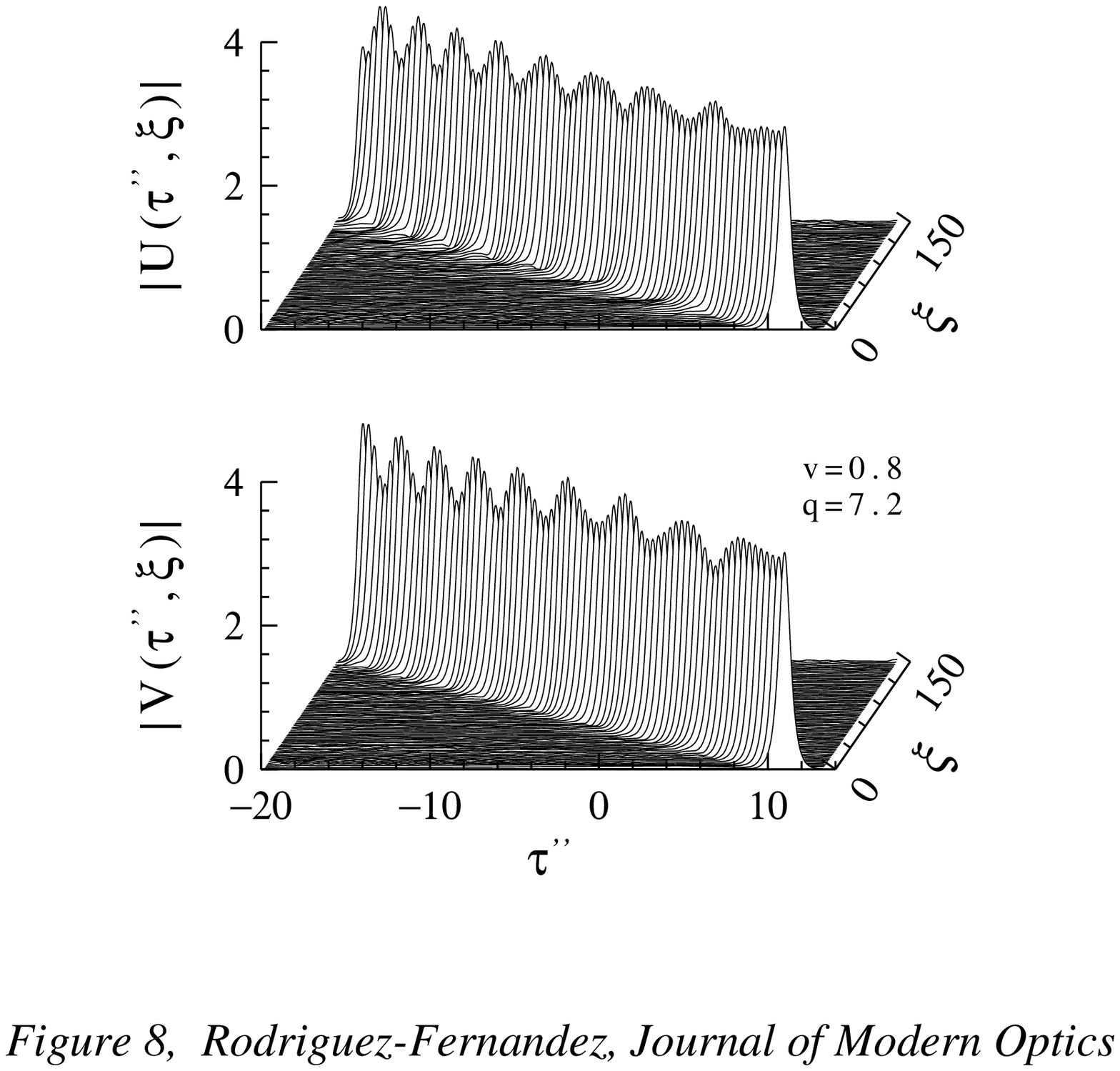} 
\caption{ Unstable propagation of a stationary solution with
its components in quadrature.  $\beta=\delta=1$, $q=7.2$, and $v=0.8$ }
\label{ab8}
\end{figure}

We have numerically solved Eqs.\ref{propa}, taking as initial
conditions the variational solutions corresponding
to different values of
the parameters $v$ and $q$ and fixed values of
$\delta$ and $\beta$, viz.
$\beta=1=\delta$. In all cases that we propagate
in-phase solutions, we obtained stable propagation, whilst
$\pi/2$-dephased solutions were sometimes unstable.
Fig.~\ref{ab6} shows the stable propagation of the
in-phase variational solution for
$q=3.2$ and $v=0.9$, whereas
Fig.\ref{ab7} shows the same for
the $\pi/2$-dephased solution corresponding to $q=6$ and $v=0.9$.
Contrary to the previous case, a small difference can be
appreciated between the variational solution  and the stationary one.
The exact stable stationary solution  (which of course is $\pi/2$-dephased) 
is reached after a very short propagation distance.
Finally Fig.~\ref{ab8} illustrates the unstable behavior
of a $\pi/2$-dephased solution ($q=7.2$, $v=0.8$). The solution
oscillates around the stationary solution with increasing
amplitude emitting a small quantity of radiation. 
Eventually, after the emisssion of a  large quantity of
radiation, a stable solution is reached (not shown in the figure).
In Figs.~\ref{ab6}-\ref{ab8} $\tau''=\tau+\delta$.

\section{Summary}
\label{sum}
	We have  developed an accurate
 variational approach to derive analytical approximations
of the coupled pulselike stationary 
solutions in birefringent
fibers in the most general case when   differences in
both phase velocities and group velocities between the two components
and fastly oscillatory
terms are taken into account (they are
important in the subpicosecond regime).  As particular
cases it includes  those where any term  in Eqs.\ref{propa}
can be neglected.  We have shown that the difference between
the central frequencies of the components, their widht and their energies
are almost independent on $v$, being mainly determined by $q$.
 We have also shown that the ratio $a_1/a_2$ is mainly determined by $v$.
The stability of these
solutions has been briefly considered, by numerically
propagating them. In all the cases we propagated
solution in-phase, they happened to be stable whilst those
$\pi/2$ out of phase present 
some intervals of stability. A global stability
analysis of these solutions remains
to be done. 
This variational method could be useful to design
soliton-dragging logic gates \cite{islam}.

{\bf  Acknowledgments}

This  work  was supported
by the Comunidad de Madrid under contract 06T/039/96 and
by the CICyT under contract TIC95-0563-03. N.J.R-F acknowledges
a grant from the CSIC.

\newpage

\end{document}